\documentclass[useAMS,usenatbib,usegraphicx,onecolumn]{mn2e}

\title[Dynamics of dark matter ellipsoids]
  {Approximate dynamics of dark matter ellipsoids}
\author[G. S. Bisnovatyi-Kogan and O. Yu. Tsupko]{G. S. Bisnovatyi-Kogan$^{1,2,3}$\thanks{E-mail:
gkogan@iki.rssi.ru (GSBK); tsupko@iki.rssi.ru (OYuT)} and O. Yu.
Tsupko$^{1,3}$\footnotemark[1]\\
$^{1}$Space Research Institute of Russian Academy of Science, Profsoyuznaya 84/32, Moscow 117997, Russia\\
$^{2}$Joint Institute Nuclear Research, Dubna, Russia\\
$^{3}$Moscow Engineering Physics Institute, Moscow, Russia}

\date{Accepted 2005 September 12. Received 2005 September 9; in original form 2005 May 18}

\pagerange{\pageref{firstpage}--\pageref{lastpage}} \pubyear{2005}

\def\LaTeX{L\kern-.36em\raise.3ex\hbox{a}\kern-.15em
    T\kern-.1667em\lower.7ex\hbox{E}\kern-.125emX}

\begin{document}

\label{firstpage}

\maketitle

\begin{abstract}
 The collapse of a non-collisional dark matter and the formation of pancake
structures in the universe are investigated approximately.
Collapse is described by a system of ordinary differential
equations, in the model of a uniformly rotating, three-axis,
uniform density ellipsoid. Violent relaxation, mass, and angular
momentum losses are taken into account phenomenologically. The
formation of the equilibrium configuration, secular instability
and the transition from a spheroid to a three-axis ellipsoid are
investigated numerically and analytically in this dynamical model.
\end{abstract}

\begin{keywords}
 dark matter  -- large-scale structure of Universe.
\end{keywords}

\section{Introduction}

According to modern cosmological ideas, most of the matter in the
Universe is in the form of so-called cold dark matter, consisting
of non-relativistic particles. The study of the formation of dark
matter objects in the Universe is based on N-body si\-mulations,
which are very time consuming. In this situation, a simplified
approach may become useful, because it allows one to investigate
many different variants and to obtain some new principally
important features of the problem, which could be lost or not
visible during long numerical work.

The systematic analysis of ellipsoidal figures of equilibrium is
presented in the classical book of \citet{cha}. Different types of
incompressible ellipsoids are investigated there on the basis of
virial equations, as well as their dynamical and secular
stability. In particular, the points of onset of the bar-mode
dynamical and secular instabilities of the Maclaurin spheroid, for
its transition to the Jacobi ellipsoid, are found. \citet{LB64,
LB65} applied the Chandrasechar's virial tensor method for a
rotating, self-gravitating spheroid of pressure-free gas and
showed the growth of non-axisymmetric perturbations during
collapse. It was also discussed that the slowly shrinking
Maclaurin spheroid will enter the Jacobi series if it shrinks
slowly enough for the dissipative mechanisms to be operative.
There have been numerical investigations of collapsing
pressureless spheroids in the papers of \citet{LB64} and
\citet{LMS}. The dynamics of a non-rotating sphere in two
dimensions was considered in \citet{LB79}, and it was shown that
the pressure prevents the development of large-scale shape
instability if initially the gravity is more than three-fifths
pressure resisted. There is a wide-ranging review of this topic in
the paper of \citet{LB96} in memory of Chandrasekhar.

Subsequent investigations of rapidly rotating figures are
connected with stars, and with large-scale structure of the
Universe. In \citet{RH91} the virial equations for rotating
Riemann ellipsoids of incompressible fluid are demonstrated to
form a Hamiltonian dynamical system. There is a detailed
description of the ellipsoid model of rotating stars in the papers
of \citet*{lrs1, lrs4, lrs5}. Using a variational principle, they
derived and investigated the equations for the evolution of a
compressible Riemann-S ellipsoid, incorporating viscous
dissipation and gravitational radiation. The solutions of these
approximate equations permitted them to obtain equilibrium models,
and to investigate their dynamical and secular stability. In a
recent paper \citep{shap04} secular bar-mode instability driven by
viscous dissipation was investigated, using these equations, and
the point of instability of compressible Maclaurin spheroids was
found.

The modern theory of a large-scale structure is based on the ideas
of \citet{z70}, concerning the formation of strongly non-spherical
structures during non-linear stages of the development of
gravitational instability, known as `Zel'dovich's pancakes'.
Numerical simulations for these objects have been performed
subsequently by many groups. In the case of structures in dark
matter, we are dealing with non-collisional non-relativistic
particles, interacting only by gravitation. The development of
gravitational instability and collapse in dark matter do not lead
to any shock formation or radiation losses, but are characterized
by non-collisional relaxation. This relaxation is based on the
idea of a `violent relaxation' of \citet{ref2}.

Here we derive and solve the equations for the dynamical behaviour
in a simple model of a compressible uniformly rotating ellipsoid.
We derive the equations for axes with the help of variation of the
Lagrange function of an ellipsoid. Correct description of pressure
effects, attained by such an approach, and the addition of
relaxation permit us to obtain the dynamics of motion without any
numerical singularities. In our model, motion along three axes
takes place in the gravitational field of a uniform density
ellipsoid, with account of the isotropic pressure, represented in
an approximate non-differential way. Relaxation leads to a
transformation of the kinetic energy of ordered motion into
kinetic energy of chaotic motion, and to increases in the
effective pressure and thermal energy. All losses are connected
with runaway particles. The collapse of the rotating three-axis
ellipsoid is approximated by a system of ordinary differential
equations, where relaxation and losses of energy, mass and angular
momentum are taken into account phenomenologically. The system is
solved numerically for several parameters, characte\-rizing the
configuration. The approach in this work is similar to that used
in \citet{ref1}, where only dark matter spheroids ($a = b \neq c$)
were considered. In this case there are analytical formulae for
the gravitational potential and forces. In the description of
violent relaxation and different kinds of losses, we use the same
approach as \citet{ref1}. In the present paper we also find the
point of the onset of secular instability of a compressible
Maclaurin spheroid, using the derived dynamical equations, and
also from the analysis of the sequence of equilibrium
configurations. A simple analytical formula for the point of onset
of the bar-mode secular instability of the Maclaurin spheroid is
found.

\section{Equations of motion}

Let us consider a three-axis ellipsoid, consisting of
non-collisi\-onal, non-relativistic particles, with semi-axes $a
\neq b \neq c$
\begin{equation}
\label{ellipsoid}\frac{x^2}{a^2}+\frac{y^2}{b^2}+\frac{z^2}{c^2}=1,
\end{equation}
and rotating uniformly with angular velocity $\Omega$ around the
$z$-axis. Let us approximate the density of the matter, $\rho$, in
the ellipsoid as uniform.

The mass $m$ and total angular momentum $M$ of the uniform
ellipsoid are connected with uniform density, angular velocity and
semi-axes as ($V$ is the volume of the ellipsoid)
\begin{equation}
\label{mass-moment}m = \rho \, V = \frac{4\pi}{3} \, \rho \, abc
\, , \quad M = \rho\,\Omega\int\limits_V (x^2+y^2) \, dV =
\frac{m}{5} \, \Omega (a^2+b^2) \, .
\end{equation}
We assume a linear dependence of the velocities on the coordinates
in the rotating frame,
\begin{equation}
\label{linear-dep} \upsilon_x =\frac{\dot{a}x}{a} \: , \quad
\upsilon_y =\frac{\dot{b}y}{b} \: , \quad \upsilon_z
=\frac{\dot{c}z}{c} \, .
\end{equation}
The gravitational energy of the uniform ellipsoid is defined as
\citep{ll93}
\begin{equation}
\label{grav-energy}
U_g=-\frac{3Gm^2}{10}\int\limits_0^{\infty}\frac{du}{\sqrt{(a^2+u)(b^2+u)(c^2+u)}},
\end{equation}
and is expressed in elliptical integrals \citep{gr62}. Let us
consider a compressible ellipsoid with a constant mass and angular
momentum, a total thermal energy of non-relativistic dark matter
particles $E_{th} \sim V^{-2/3} \sim (abc)^{-2/3}$ and the
relation between pressure $P$ and thermal energy $E_{th}$ as
$E_{th} = \frac{3}{2} P V$. In absence of any dissipation, this
ellipsoid is a conservative system. To derive the equations of
motion, let us write the Lagrange function of the ellipsoid:
\begin{equation}
\label{Lagr-funct} L = U_{kin} - U_{pot}\, , \quad U_{pot} = U_g +
E_{th} + U_{rot} \: ,
\end{equation}
\begin{equation}
\label{U-kinE-th} U_{kin} = \frac{1}{2} \,
\rho\int\limits_V(\upsilon_x^2+\upsilon_y^2+\upsilon_z^2)\,dV
 = \frac{m}{10} \,
(\dot{a}^2+\dot{b}^2+\dot{c}^2) \: , \quad E_{th} =
\frac{E_{th,in} (a_{in}b_{in}c_{in})^{2/3}} {(abc)^{2/3}} = \frac
{\varepsilon} {(abc)^{2/3}} \: ,
\end{equation}
where the entropy function $\varepsilon$ is constant in the
conservative case, but variable in the presence of dissipation,
and
\begin{equation}
\label{U-rot-omega} U_{rot} = \frac{1}{2}\,\rho\int \limits_V
V_{rot}^2 \,dV = \frac{1}{2}\,\rho\,\Omega^2 \int \limits_V
(x^2+y^2) \,dx\,dy\,dz = \frac{m}{10}\,\Omega^2 (a^2+b^2) \: .
\end{equation}
Taking into account the second expression in (\ref{mass-moment}),
we obtain the relation
\begin{equation}
\label{U-rot-M} U_{rot} = \frac{5}{2} \, \frac{M^2}{m(a^2+b^2)} \:
.
\end{equation}
By variation of the Lagrange function we obtain Lagrange equations
of motion:
\begin{equation}
\ddot{a} = - \frac{3Gm}{2} \, a
\int\limits_0^{\infty}\frac{du}{(a^2+u)\Delta} \,\, +
\,\frac{10}{3m}\, \frac{1}{a} \,\frac {\varepsilon} {(abc)^{2/3}}
\,\,+ \frac{25 M^2}{m^2} \, \frac{a}{(a^2+b^2)^2} \, ,
\end{equation}
\begin{equation}
\ddot{b} = - \frac{3Gm}{2} \, b
\int\limits_0^{\infty}\frac{du}{(b^2+u)\Delta} \,\,+
\,\frac{10}{3m}\, \frac{1}{b} \,\frac {\varepsilon} {(abc)^{2/3}}
\,\, + \frac{25 M^2}{m^2} \, \frac{b}{(a^2+b^2)^2} \, ,
\end{equation}
\begin{equation}
\ddot{c} = - \frac{3Gm}{2} \, c
\int\limits_0^{\infty}\frac{du}{(c^2+u)\Delta} \,\,+
\,\frac{10}{3m}\, \frac{1}{c} \,\frac {\varepsilon} {(abc)^{2/3}}
\; , \; \; \Delta^2 = (a^2+u)(b^2+u)(c^2+u) \, .
\end{equation}

It is easy to check that the equilibrium solution of these
equations are the Maclaurin sphe\-roid and the Jacobi ellipsoid
\citep{cha}.

\section{Equations of motion with dissipation}
In reality there is relaxation in the collisionless system,
connected with phase mixing, which is called `violent
relaxa\-tion' \citep{ref2}. This relaxation leads to a
transformation of the kinetic energy of ordered motion into
kinetic energy of chaotic motion and increases the effective
pressure and thermal energy.

The main transport process is an effective bulk viscosity.
Therefore, there is a drag force, which is described
phenomenologically by adding the terms
\begin{equation}
- \frac{\dot{a}}{\tau_{rel}} \: , \quad -
\frac{\dot{b}}{\tau_{rel}} \: , \quad - \frac{\dot{c}}{\tau_{rel}}
\end{equation}
on the right-hand sides of the equations of motions.

Here we have scaled the relaxation time $\tau_{rel}$ by the Jeans
characteristic time
\begin{equation}
\tau_J = \frac{2 \pi}{\omega_J} = \frac{2 \pi}{\sqrt{4 \pi Gm}} =
2 \pi \sqrt{\frac{abc}{3Gm}}
\end{equation}
with a constant value of $\alpha_{rel}$,
\begin{equation}
\label{tau-rel} \tau_{rel} = 2 \pi \, \alpha_{rel}
\sqrt{\frac{abc}{3Gm}} \, .
\end{equation}

The process of relaxation is also accompanied by total energy
$U_{tot}$, mass and angular momentum losses from the system. These
losses may be described phenomenologically by characteristic times
$\tau_{el}$, $\tau_{ml}$ and $\tau_{Ml}$, scaled by $\tau_J$ with
the coefficients $\alpha_{el}$, $\alpha_{ml}$ and $\alpha_{Ml}$:

\[
U_{tot}=U_{kin}+U_g+U_{rot}+E_{th}.
\]
The corresponding equations describing the various losses are
derived for a three-axis ellipsoid in section 4 of the paper by
\citet{ref1}.

Using the variational principle and deriving the equation for the
entropy function from energy balance give the correct relations
for the pressure and the total energy. \citet{Fuji} derived the
equations of motion for a rotating ellipsoid from a system of
hydrodynamical equations. However, the accounting for pressure
effects in his approach was inconsistent with thermal processes,
leading to the wrong results for the dynamical behaviour of a
system with radiative losses (compare with \citealt{lrs4} and
\citealt{ref1}).

\section{Non-dimensional equations}
To obtain a numerical solution of the equations, we write them in
non-dimensional variables. Let us introduce the variables
\[
\tilde{t} = \frac{t}{t_0} \: , \quad \tilde{a} = \frac{a}{a_0} \:
, \quad \tilde{b} = \frac{b}{a_0} \: , \quad \tilde{c} =
\frac{c}{a_0} \: , \quad \tilde{m} = \frac{m}{m_0} \: , \quad
\tilde{M} = \frac{M}{M_0} \: , \quad \tilde{\rho} =
\frac{\rho}{\rho_0} \: , \quad \tilde{U} = \frac{U}{U_0} \: ,
\quad \tilde{E}_{th} = \frac{E_{th}}{U_0} \: , \quad
\tilde{\varepsilon} = \frac{\varepsilon}{\varepsilon_0} \: , \quad
\tilde{\tau_i} = \frac{\tau_i}{t_0} \: .
\]
The scaling parameters $\, t_0, \; a_0, \; m_0, \; M_0, \; \rho_0,
\; U_0, \; \Omega_0$ and $\varepsilon_0$ are connected by the
following relations:
\[
t_0^2 = \frac{a_0^3}{G m_0}, \quad M_0^2 = G a_0 m_0^3, \quad U_0
= \frac{G m_0^2}{a_0}, \quad \rho_0 = \frac{m_0}{a_0^3}, \quad
\Omega_0 = \frac{M_0}{m_0 a_0^2}, \quad \varepsilon_0 = U_0 a_0^2.
\]
The parameter $U_0$ is used for scaling of all kinds of energies.
Hereafter the `tilde' is omitted for simplicity. In
non-dimensional variables, we have
\[
m = \frac{4\pi}{3} \, \rho \, abc, \quad M = \frac{m}{5} \, \Omega
(a^2+b^2), \quad \tau_i = 2 \pi \, \alpha_i \sqrt{\frac{abc}{3m}}.
\]
Taking into account violent relaxation, total energy, mass and
angular momentum losses, the dynamics of the system is described
by the following non-dimensional system of equations:
\begin{equation}
\label{ddota}
  \ddot{a} = - \frac{\dot{a}}{m} \frac{dm}{dt} -
\frac{3m}{2} \, a \int\limits_0^{\infty}\frac{du}{(a^2+u)\Delta}\,
+ \,\frac{10}{3m}\, \frac{1}{a} \,\frac {\varepsilon}
{(abc)^{2/3}} \, + \frac{25 M^2}{m^2} \, \frac{a}{(a^2+b^2)^2} \,
- \,\frac{\dot{a}}{\tau_{rel}},
\end{equation}

\begin{equation}
\label{ddotb}
 \ddot{b} = - \frac{\dot{b}}{m} \frac{dm}{dt} - \frac{3m}{2} \, b
\int\limits_0^{\infty}\frac{du}{(b^2+u)\Delta} \,\,+
\,\frac{10}{3m}\, \frac{1}{b} \,\frac {\varepsilon} {(abc)^{2/3}}
\,\, + \frac{25 M^2}{m^2} \, \frac{b}{(a^2+b^2)^2} -
\,\frac{\dot{b}}{\tau_{rel}} \, ,
\end{equation}

\begin{equation}
\label{ddotc}
 \ddot{c} = - \frac{\dot{c}}{m} \frac{dm}{dt}
 - \frac{3m}{2} \, c
\int\limits_0^{\infty}\frac{du}{(c^2+u)\Delta} \,\,+
\,\frac{10}{3m}\, \frac{1}{c} \,\frac {\varepsilon} {(abc)^{2/3}}
- \,\frac{\dot{c}}{\tau_{rel}} \, ,
\end{equation}

\begin{equation}
\label{ddote}
 \dot{\varepsilon} = (abc)^{2/3}\, U_{kin} \left[\left(
\frac{2}{\tau_{rel}} - \frac{1}{\tau_{el}} - \frac{2}{\tau_{ml}}
\right) - \frac{U_{rot}}{U_g} \left(\frac{2}{\tau_{Ml}} -
\frac{1}{\tau_{ml}} \right) + \frac{U_{kin}}{U_g \,
\tau_{ml}}\right] ,
\end{equation}

\begin{equation}
\label{ddotm}
 \dot{m} = - \frac{1}{3\tau_{ml}} \,
(\dot{a}^2+\dot{b}^2+\dot{c}^2) /
\left(\int\limits_0^{\infty}\frac{du}{\Delta}\right) , \; \;
\frac{M}{M_{in}} =
\left(\frac{m}{m_{in}}\right)^{\frac{\tau_{ml}}{\tau_{Ml}}} \; ,
\; \; \Delta^2 = (a^2+u)(b^2+u)(c^2+u) \, .
\end{equation}

This system is solved numerically for several initial parameters.

\section{Numerical results}

The integrals describing gravitational potentials and forces, are
expressed in elliptical integrals \citep{gr62}. For ellipsoids
that are close to a spheroid or to a sphere, we have carried out
an expansion to linear terms near points $a = b \neq c$, $b = c
\neq a$, $c = a \neq b$ and $a = b = c$, respectively. These
expansions include only analytical expressions. All integrals
reduce to two, $I0$ and $I1$, i.e.

\begin{equation}
\label{i0-i1}
 I0 =
\int\limits_0^{\infty}\frac{du}{\sqrt{(a^2+u)(b^2+u)(c^2+u)}} \, ,
\quad
 I1 =
\int\limits_0^{\infty}\frac{du}{(a^2+u)\sqrt{(a^2+u)(b^2+u)(c^2+u)}}
\, .
\end{equation}
In the limiting cases these integrals are expressed analytically
as follows:

\[
I0= 2 {\frac {\arccos \left( c/a \right)}{\sqrt {{a}^{2}-{c}
^{2}}}} -\frac{1}{2} \, \left( {b}^{2}-{a}^{2} \right)  \left[
\frac{\arccos \left( c/a \right)}{\left( {a}^{2}-{c}^{2} \right)
^{3/2}}+{\frac { c}{ \left( {c}^{2}-{a}^{2} \right) {a}^{2}}}
\right],
\]
\[
I1= - \left( {\frac {c}{a^2}}-\frac{\arccos \left( c/a \right)}{
\sqrt{a^2-c^2} } \right) \frac{1}{a^2-c^2} \, - \, \frac{1}{2} \,
\left( {b}^{2}-{a}^{2 } \right) \left[ -\frac{3}{4} \, {\frac
{c}{{a}^{2} \left( {a}^{2}-{c}^{2}
 \right) ^{2}}}\, - \, \frac{1}{2} \,{\frac {c}{{a}^{4} \left( {a}^{2}-{c}^{2}
 \right) }} + \frac{3}{4} \, \frac{\arccos \left( c/a \right)}{\left( {a}^{2}
-{c}^{2} \right) ^{5/2} } \right]
\]
at
\begin{equation}
\label{i10}
 \left| {\frac {b}{a}}-1 \right| < 0.0001, \quad a,b>c, \quad \left|
{\frac {c}{a}}-1 \right| \geq 0.0001 \, ;
\end{equation}
\[
I0= 2 {\frac{\cosh^{-1} \left( c/a \right)}{\sqrt {{c}^{
2}-{a}^{2}}}} -\frac{1}{2} \, \left( {b}^{2}-{a}^{2} \right)
\left[ - \frac{\cosh^{-1} \left( c/a \right)}{\left(
{c}^{2}-{a}^{2}
 \right)^{3/2}} + {\frac {c}{ \left( {c}^{2}-{a}^{2} \right) {a}^{2}}}
 \right],
\]
\[
I1= - \left( {\frac {c}{a^2}}-\frac{\cosh^{-1} \left( c/a
\right)}{ \sqrt{c^2-a^2} } \right) \frac{1}{a^2-c^2} \,
 - \, \frac{1}{2} \, \left( {b}^{2 }-{a}^{2} \right) \left[
-\frac{3}{4} \, {\frac {c}{{a}^{2} \left( {a}^{2}-{c}^{2}
 \right) ^{2}}}\, - \, \frac{1}{2} \,{\frac {c}{{a}^{4} \left( {a}^{2}-{c}^{2}
 \right) }} + \frac{3}{4} \, \frac{\cosh^{-1} \left( c/a \right)}{\left( {
c}^{2}-{a}^{2} \right) ^{5/2}} \right]
\]
at
\begin{equation}
\label{i2}
 \left| {\frac {b}{a}}-1 \right| < 0.0001, \quad
a,b<c, \quad \left| {\frac {c}{a}}-1 \right| \geq 0.0001 \, ;
\end{equation}

\[
I1= 2 \, \left( \frac{1}{a} - \frac{\arccos \left( a/b
\right)}{\sqrt{b^2-a^2}}  \right)
 \frac{1}{b^2-a^2} \,
 - \, \frac{1}{2} \, \left( {c}^{2
}-{b}^{2} \right)  \left[ {\frac {a}{{b}^{2} \left(
{b}^{2}-{a}^{2}
 \right) ^{2}}} + 2 \,{\frac {1}{a \left(
{b}^{2}-{a}^{2}
 \right) ^{2}}} - 3\,\frac{\arccos \left( a/b \right)}{\left( {b}^{
2}-{a}^{2} \right)^{5/2}}  \right]
\]
at
\begin{equation}
\label{i3}
 \left| {\frac {b}{c}}-1 \right| < 0.0001, \quad b,c>a, \quad \left|
{\frac {b}{a}}-1 \right| \geq 0.0001 \, ;
\end{equation}

\[
I1= 2 \, \left( \frac{1}{a} - \frac{\cosh^{-1} \left( a/b
\right)}{\sqrt{a^2-b^2}}  \right)
 \frac{1}{b^2-a^2} \,
 - \, \frac{1}{2} \, \left( {c}^{2
}-{b}^{2} \right)  \left[ {\frac {a}{{b}^{2} \left(
{b}^{2}-{a}^{2}
 \right) ^{2}}} + 2 \,{\frac {1}{a \left(
{b}^{2}-{a}^{2}
 \right) ^{2}}} - 3\, \frac{\cosh^{-1} \left( a/b
\right)}{\left( {a}^{2}-{b}^{2 } \right) ^{5/2}}
 \right]
\]
at
\begin{equation}
\label{i4}
 \left| {\frac {b}{c}}-1 \right| < 0.0001, \quad
b,c<a, \quad \left| {\frac {b}{a}}-1 \right| \geq 0.0001 \, ;
\end{equation}

\[
I0= \frac{2}{a} + \frac{1}{3}\,{\frac {{a}^{2}-{c}^{2}}{{a}^{3}}}
+ \frac{1}{3}\,{\frac {{a}^{ 2}-{b}^{2}}{{a}^{3}}},\quad
 I1= \frac{2}{3
\,a^3} + \frac{1}{5} \, {\frac {{a}^{2}-{c}^{2}}{{a}^{5}}} +
\frac{1}{5} \, {\frac {{a} ^{2}-{b}^{2}}{{a}^{5}}}
\]
at
\begin{equation}
\label{i7} \left| {\frac {b}{a}}-1 \right| < 0.0001, \quad \left|
{ \frac {c}{a}}-1 \right| < 0.0001.
\end{equation}
Corresponding relations for other cases are obtained from
equations (\ref{i10}),(\ref{i2}),(\ref{i3}),(\ref{i4}) by cyclic
permutation\footnote{Note that expansions given in (42) of
\citet{ref1} contain misprints. The correct expansions at
$|k-1|\ll 1$ are
\[
a_k=1+\frac{1-k}{3};\quad a1_k=-\frac{2}{3}-\frac{2}{5}(1-k);\quad
a2_k=\frac{1}{3}+\frac{4}{15}(1-k).
\]
These relations are valid both for $k>1$ and $k<1$.}. In the
calculation we have used the Runge-Kutta method. The accuracy of
the calculation is really very good, especially bearing in mind
the relative simplicity of the equations. The integration was
performed with relative precision of $10^{-5}$, and the
corresponding variant with the precision of $10^{-7}$ was exactly
the same. One plot contains several thousands of time-steps.
Detailed information about the initial parameters for all variants
of the calculation is given in Table 1.

\begin{table}
\caption{Initial values and parameters of the calculated
variants.}
\begin{tabular}{@{}ccccc} \hline
  Cases  & 1       & 2 & 3 & 4  \\
  \hline
  Figure & 1,2,3,4 & 5 & 6 & 7  \\
  \hline
  $ a $ & 1 & 1 & 1 & 1   \\
  $ b $ & 1 & 1 & 1 + $10^{-5}$& 1 + $10^{-5}$   \\
  $ c $ & 1 & 1 & 1 + $2 \times 10^{-5}$ & 0.5   \\
  $\dot{a}_{in}$ & 0.2 & 0.5 & 0 & 0   \\
  $\dot{b}_{in}$ & 0.2+$10^{-10}$ & 0.5+$10^{-10}$ & 0 & 0   \\
  $\dot{c}_{in}$ & 0.2 & 0.5 & 0 & 0   \\
  $ M_{in}$ & 0.5 & 0.1 & 0 & 0.317   \\
  $\varepsilon_{in}$ & 0.01 & 0 & 0.01 & 0.149  \\
  $\alpha_{rel}$ & 3 & 3 & - & -  \\
  $\alpha_{el},\alpha_{ml},\alpha_{Ml}$ & 15 & 15 & - & -  \\
  \hline
\end{tabular}

\end{table}
The case of a spheroid ($a = b \neq c$) is consi\-dered in
\citet{ref1}. Our 3D equations give the same results for
spheroidal initial conditions. In the case of a three-axis
ellipsoid, there are new qualitative effects, because there is an
additional degree of freedom compared to the case of a spheroid.
For the investigation of dynamical behaviour, we start the
simulation from a spherical body of unit mass ($m_{in} = 1$), and
zero or small entropy $\varepsilon \ll 1$. We also specify the
parameters characterizing different dissipations. In all
calculations with relaxation, we use following values:
$\alpha_{rel} = 3$ and $\alpha_{el} = \alpha_{ml} =  \alpha_{Ml} =
15$. We set non-zero initial velocities. Because of rotation
around $c$-axis, the initial sphere transforms to a spheroid
during collapse if the initial parameters for axes $a$ and $b$ are
exactly equal ($a = b, \: \dot{a} = \dot{b}$). Note that during
the motion spheroids may be not only oblate but also prolate
\citep{ref1}. To study the appearance of three-axis figures, the
initial parameters for axes $a$ and $b$ were slightly different in
all variants of the calculations.

In the first variant (see Fig. 1) there is a large initial angular
momentum $M_{in} = 0.5$. The field of velocities is slightly
perturbed by increasing $\dot{b}$ in comparison with $\dot{a}$.
Initially we observe the collapse and the formation of the
pancake. During the motion the difference $(a-b)$ increases due to
the development of the secular instability, and we obtain the
transformation of a Maclaurin sphe\-roid into a Jacobi
el\-lip\-soid in the dynamics. Because of the relaxation, the
oscillation motion damps, and the configuration reaches the
equilibrium state of the three-axis ellipsoid. The corresponding
behaviours of total energy $U_{tot}$, entropy function
$\varepsilon$ and mass $m$ are represented in Figs 2, 3 and 4. We
see that the main changes of these parameters take place at the
stages of the first few oscillations: the entropy function
$\varepsilon$ increases, the mass $m$ and the total energy
$U_{tot}$ decreases. Than these parameters reach equilibrium
values.

\begin{figure}
\includegraphics[width=0.5\textwidth]{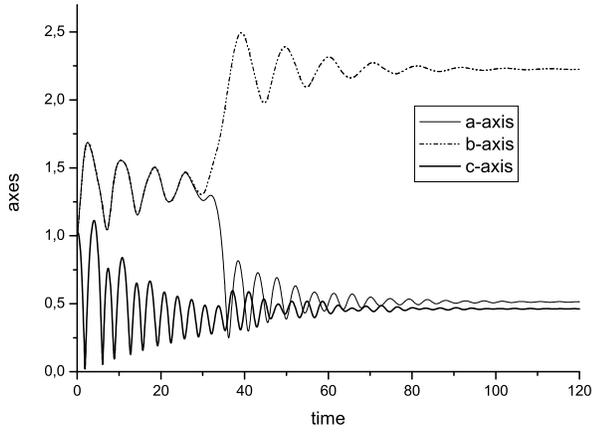}
\caption{The development of an instability at large angular
momentum, and the formation of a stationary triaxial figure in
case 1 (see Table 1 for cases).}
\end{figure}

\begin{figure}
\includegraphics[width=0.5\textwidth]{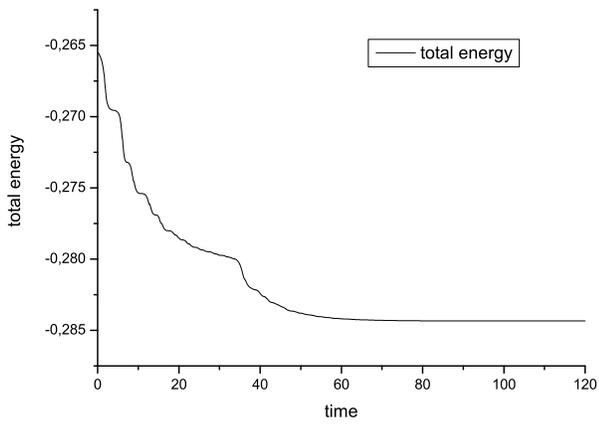}
\caption{The total energy evolution for case 1.}
\end{figure}

\begin{figure}
\includegraphics[width=0.5\textwidth]{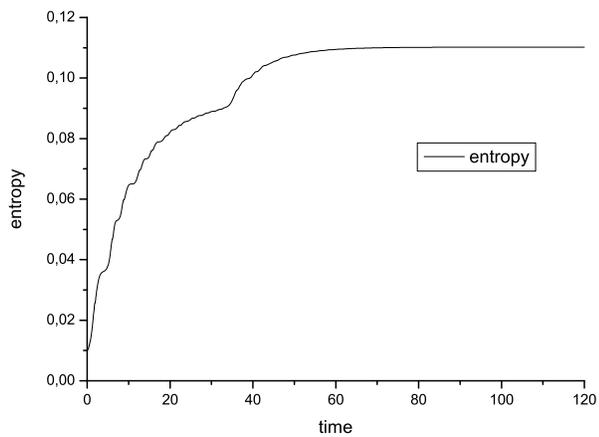}
\caption{The entropy function evolution for case 1.}
\end{figure}

\begin{figure}
\includegraphics[width=0.5\textwidth]{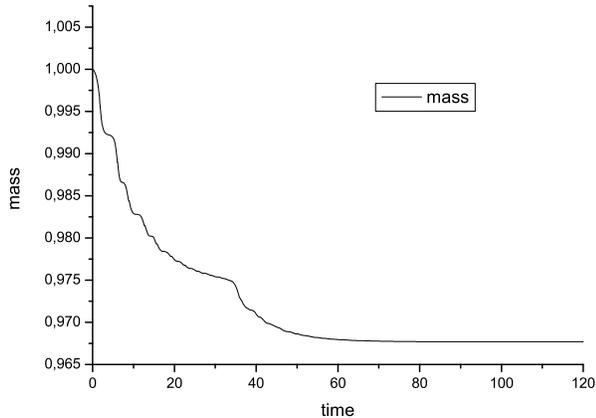}
\caption{The mass evolution for case 1.}
\end{figure}

In the second variant of the calculations (see Fig. 5) we set a
small initial angular momentum $M_{in}=0.1$. After the collapse,
temporary oblate and prolate spheroids appear during the
relaxation. There is no secular instability, and the system
rea\-ches the equilibrium oblate spheroid of Maclaurin. An initial
small difference of parameters connected with axes $a$ and $b$
leads to a small deviation from the results of the similar variant
of the 2D simulations \citep{ref1}. Here we have an additional
instability: there is an increase in the difference $(a-b)$ at the
early stages of motion. However, this difference remains too small
to be visualized in the figure. This is the instability
characteristic to the system with purely radial trajectories
\citep{ant,fp85}. This instability takes place in all variants of
the calculations, but in most cases it is very small and always
disappears during the relaxation and formation of the equilibrium
figure. In the cases of large angular momentum, this instability
does not show up because the difference $(a-b)$ quickly increases
as a result of secular instability.

To illustrate the radial instability, we consider an initial
sphere without rotation ($M_{in} = 0$) and without any relaxation
process.  Only the three equations (\ref{ddota})--(\ref{ddotc})
with $\tau_{rel}=\infty$ have been used, at constant $m=1$, $M=0$
and $\varepsilon$.  We have set small initial differences between
the axes $a$, $b$ and $c$, and have obtained an unstable behaviour
evidently visible in Fig. 6.

\begin{figure}
\includegraphics[width=0.5\textwidth]{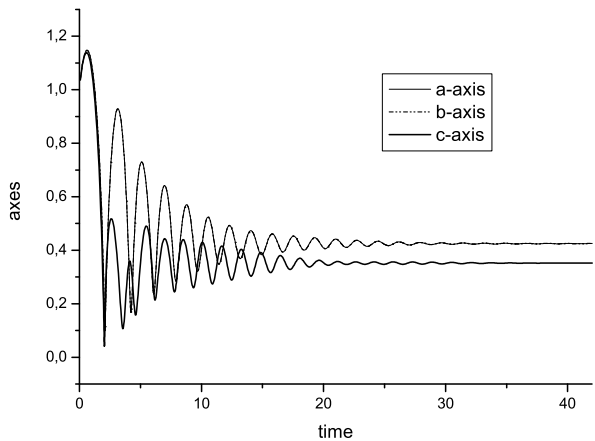}
\caption{The relaxation to the spheroid at small angular momentum,
case 2.}
\end{figure}

\begin{figure}
\includegraphics[width=0.5\textwidth]{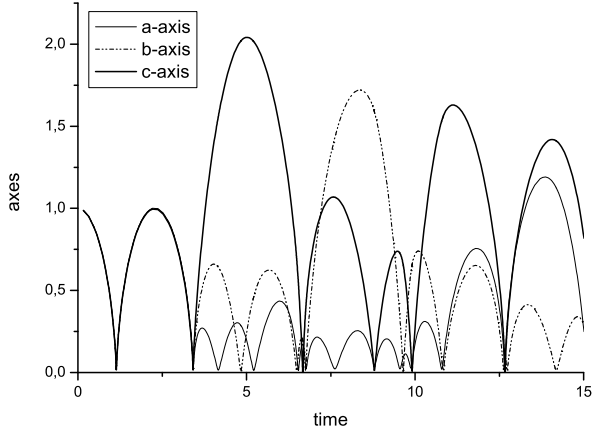}
\caption{Radial instability development in the non-rotating body,
without dissipation, case 3.}
\end{figure}

\begin{figure}
\includegraphics[width=0.5\textwidth]{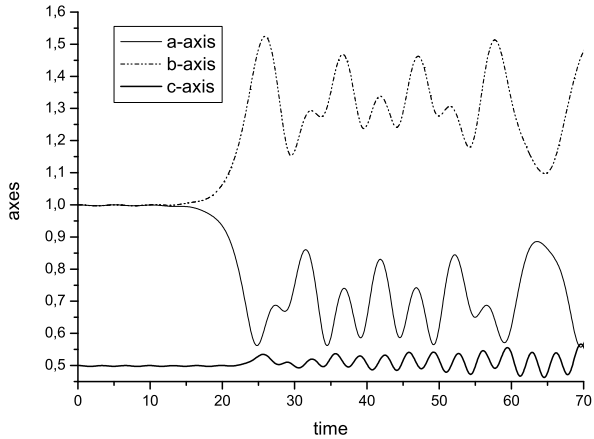}
\caption{Development of the secular instability in a rapidly
rotating body, without dissipation, case 4.}
\end{figure}

For numerical investigation of the secular instability, we
consider equations of motion without dissipation, and take the
equilibrium spheroids with small perturbation as the initial
configuration. It is easy to obtain the equilibrium parameters of
the spheroid from the equations of motion with zero accelerations.
The initial equilibrium configuration of the Maclaurin spheroid
was slightly perturbed by increasing $b$, as in the last column of
Table 1. At small angular momentum we obtain the conservation of
the initial configuration. There are small oscillations around the
equilibrium spheroid, during which the difference $(a-b)$ changes
sign periodically. At large angular momentum we obtain the
development of the secular instability, and, after several
low-amplitude oscillations, the Maclaurin sphe\-roid,  is
transformed into the Jacobi el\-lip\-soid (see Fig. 7 where the
variant with large angular momentum is represented). Because of
the absence of relaxation, the system does not reach an
equilibrium configuration and oscillations continue. If we include
relaxation in this variant, we should observe the final
equilibrium state of the ellipsoid. Our calculations allow us to
find approximately numerically the point of the onset of the
secular instability of a compressible Maclaurin sphe\-roid. It
fits well with the analytical results obtained in the next
section.

\section{Equilibrium configurations and stability}

Equilibrium of a uniformly rotating figure (spheroid or ellipsoid)
is found from the equations (\ref{ddota})--(\ref{ddotc}) with zero
time derivatives:

\begin{equation}
\label{a}
  0 = - \frac{3m}{2} \, a
\int\limits_0^{\infty}\frac{du}{(a^2+u)\Delta} \,\, +
\,\frac{10}{3m}\, \frac{1}{a} \,\frac {\varepsilon} {(abc)^{2/3}}
\,\,+ \frac{25 M^2}{m^2} \, \frac{a}{(a^2+b^2)^2} \, ,
\end{equation}
\begin{equation}
\label{b}
 0 = - \frac{3m}{2} \, b
\int\limits_0^{\infty}\frac{du}{(b^2+u)\Delta} \,\,+
\,\frac{10}{3m}\, \frac{1}{b} \,\frac {\varepsilon} {(abc)^{2/3}}
\,\, + \frac{25 M^2}{m^2} \, \frac{b}{(a^2+b^2)^2} \, ,
\end{equation}
\begin{equation}
 \label{c}
 0 = - \frac{3m}{2} \, c
\int\limits_0^{\infty}\frac{du}{(c^2+u)\Delta} \,\,+
\,\frac{10}{3m}\, \frac{1}{c} \,\frac {\varepsilon} {(abc)^{2/3}}.
\end{equation}
Introducing the variables

\begin{equation}
 \label{var}
 x=\frac{u}{a^2},\quad k=\frac{c}{a},\quad k_1=\frac{b}{a},\quad
 j=\frac{M}{m},\quad \epsilon=\frac{\varepsilon}{m},
\end{equation}
and integrals

\begin{equation}
\label{vari}
I_1(k,k_1)=\int_0^{\infty}\frac{dx}{(x+1)^{3/2}\sqrt{(k_1^2+x)(k^2+x)}}
\, , \quad
I_2(k,k_1)=\int_0^{\infty}\frac{dx}{(x+k_1^2)^{3/2}\sqrt{(1+x)(k^2+x)}}
\, ,
\end{equation}
\[
I_3(k,k_1)=\int_0^{\infty}\frac{dx}{(x+k^2)^{3/2}\sqrt{(1+x)(k_1^2+x)}}
\, ,
\]
 we can write equations (\ref{a})-(\ref{c}), using (\ref{ddotm}), in the form

\begin{equation}
\label{a1}
  0 = - \frac{3m}{2} \, a \,I_1(k,k_1) +
\,\frac{10}{3}\, \frac {\epsilon} {(kk_1)^{2/3}} \,\,+ \frac{25
j^2}{(1+k_1^2)^2} \, ,
\end{equation}
\begin{equation}
\label{b1}
 0 = - \frac{3m}{2} \, ak_1 \,I_2(k,k_1) +
\,\frac{10}{3}\, \frac {\epsilon} {k_1(kk_1)^{2/3}} \,\,+ \frac{25
j^2\, k_1}{(1+k_1^2)^2} \, ,
\end{equation}
\begin{equation}
 \label{c1}
 0 = - \frac{3m}{2} \, ak \,I_3(k,k_1) +
\,\frac{10}{3}\, \frac {\epsilon} {k(kk_1)^{2/3}}.
\end{equation}
Excluding $m$, we obtain

\begin{equation}
\label{a2}
  0 = - \frac{1}{k^2}\frac{I_1(k,k_1)}{I_3(k,k_1)} +1
+ \frac{15 j^2}{2\epsilon}\frac{(kk_1)^{2/3}}{(1+k_1^2)^2} \, ,
\quad
 0 = - \frac{k_1^2}{k^2}\frac{I_2(k,k_1)}{I_3(k,k_1)} +1
+ k_1^2\frac{15 j^2}{2\epsilon}\frac{(kk_1)^{2/3}}{(1+k_1^2)^2}.
\end{equation}
At $a=b$, $k_1=1$ equations (\ref{a2}) are identical, and
determine the equilibrium of the Maclaurin spheroid (see
\citealt{ref1}). For Jacobi ellipsoids with $a \neq b \neq c$, we
obtain the following relation between $k$ and $k_1$:
\begin{equation}
\label{a3}
 F(k,k_1)=1-\frac{1}{k_1^2} + \frac{1}{k^2}\frac{I_2(k,k_1)}{I_3(k,k_1)}
 - \frac{1}{k^2}\frac{I_1(k,k_1)}{I_3(k,k_1)}=0 .
\end{equation}
This equation has a trivial solution $k_1=1$ at all $k$,
corresponding to the Maclairin spheroid. Let us find the
bifurcation point of the equation (\ref{a3}), at which non-trivial
solutions appear. While $k_1=1$ is always a root of equation
(\ref{a3}), we may write $ F(k,k_1)=(k_1-1)f(k,k_1)$. An
additional root of the equation $F(k,k_1)=0$ appears when the root
of the equation $f(k,k_1)=0$ appears at $k_1=1$. The root of the
zero derivative equation
$F'_{k_1}(k,k_1)=f(k,k_1)+(k_1-1)f'_{k_1}(k,k_1)=0$ at $k_1=1$
coincides with the root of the equation $f(k,1)=0$.\footnote{We
are grateful to A.I.Neishtadt for useful discussion of this point}
Therefore the value of $k$ at the bifurcation point is determined
by the equation
\begin{equation}
\label{b3}
 \frac{\partial F(k,k_1)}{\partial k_1}|_{k_1=1}=0.
\end{equation}
Using (\ref{vari}) and (\ref{a3}), this equation is written in the
form

\begin{equation}
\label{b4}
 2k^2\,I_3(k,k_1)|_{k_1=1}+\frac{\partial I_2(k,k_1)}{\partial
 k_1}|_{k_1=1}-\frac{\partial I_1(k,k_1)}{\partial k_1}|_{k_1=1}=0
\end{equation}
At $k_1=1$ we have analytic expressions

\begin{equation}
\label{b5}
 \frac{\partial I_1(k,k_1)}{\partial k_1}|_{k_1=1}=-I_0,\quad
 \frac{\partial I_2(k,k_1)}{\partial k_1}|_{k_1=1}=-3I_0,\quad
 I_0=\int_0^{\infty}\frac{dx}{(1+x)^3\sqrt{k^2+x}} \, ,
\end{equation}

so that the equation (\ref{b4}) is reduced to

\begin{equation}
\label{b6}
 k^2\,I_3=I_0,\quad {\rm where}\,\,\, I_3=I_3(k,1)\quad {\rm and}\,\,\, I_2=I_2(k,1)=
 I_1=I_1(k,1).
\end{equation}
Taking the integrals $I_0,\,\, I_1=I_2$ and $I_3$ analytically, we
have

\begin{equation}
\label{b7}
 (1-k^2)I_0=\frac{3}{4}I_2-\frac{k}{2} \, ,\quad
 I_3=\frac{2}{k(1-k^2)}\biggl(1-\frac{k\arccos{k}}{\sqrt{1-k^2}}\biggr),
 \,
 \quad
 I_1=I_2=-\frac{1}{k(1-k^2)}\biggl(k^2-\frac{k\arccos{k}}{\sqrt{1-k^2}}\biggr).
\end{equation}
Using (\ref{b7}) in (\ref{b6}) we obtain the equation

\begin{equation}
\label{b9} \frac{\arccos k}{\sqrt{1-k^2}} =
\frac{k(13-10k^2)}{3+8k^2-8k^4} \, ,
\end{equation}
for which the solution $k=0.582724$ determines the bifurcation
point at the sequence of the Maclaurin spheroids. For the uniform
spheroid, the position of this point does not depend on the
adiabatic index of the matter.

Above, we have obtained the bifurcation point on the equilibrium
curve of the Maclaurin spheroids using only the equilibrium
relations for the Jacobi ellipsoids. The usual way of
investigating stability is to solve linearized equations of motion
and to find the eigenfrequencies. The linearized equations
(\ref{ddota})--(\ref{ddotc}) without violent relaxation reduce,
with account of (\ref{var}), to the following form:

\[
 \delta(\ddot{a}-\ddot{b}) = 3\frac{m}{a^3} \,\left[ -
k^2 \int\limits_0^{\infty} \frac{du}{(1+u)(k^2+u)^{3/2}} \, +
\int\limits_0^{\infty} \frac{du}{(1+u)^3(k^2+u)^{1/2}} \right]
\delta(a-b) \, .
\]
The $\varepsilon$, $U_{tot}$, $m$ and $M$ losses are quadratic to
perturbations, so these values remain constant in linear
approximation. Taking $\delta(a-b) \sim \exp(-i \omega t)$, we
come to the characteristic equation

\begin{equation}
\label{b10}
 \omega^2 = 3\frac{m}{a^3} \,\left[k^2 \int\limits_0^{\infty}
\frac{du}{(1+u)(k^2+u)^{3/2}} \, -  \int\limits_0^{\infty}
\frac{du}{(1+u)^3(k^2+u)^{1/2}} \,\right].
\end{equation}
The plot $\omega^2(k)$ at $m=a=1$ is given in Fig. 8. Using the
notations of (\ref{b5}) and (\ref{b6}) in (\ref{b10}), we write
the characteristic equation in the form

\begin{equation}
\label{b11}
 \omega^2 = 3\frac{m}{a^3} \,(k^2\,I_3-I_0).
\end{equation}

\begin{figure}
\includegraphics[width=0.5\textwidth]{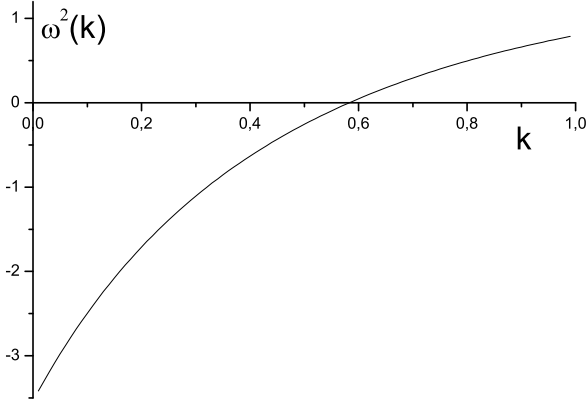}
\caption{Dependence of the squared eigenfrequency $\omega^2$ on
the axis ratio $k$ in the Maclaurin spheroid without dissipation
terms in the approximate equations of motion.}
\end{figure}

We see that a spheroid loses its stability for the transformation
into a three-axial ellipsoid, at the bifurcation point
$k=0.58272$, $e=\sqrt{1 - c^2/a^2} = 0.81267$.  By derivation it
is clear that the position of this point does not depend on the
polytropic exponent $n$. Our approximate equations, even without
dissipation, describe uniformly rotating ellipsoids (spheroids),
which are not connected by adiabatic relations, and contain a
"hidden" non-conservation of the local angular momentum, which
preserves the uniformity. Therefore, in the presence of this
"hidden" non-conservation, the loss of stability takes place
exactly at the bifurcation point. The loss of stability is not
connected with relaxation and follows from our equations even in
the case when all the global integrals remain constant. In the
exact approach, the instability at this point happens only in the
direct presence of dissipative terms \citep{cha}. The pure
adiabatic spheroidal system preserves stability until
$e=0.952887$, where it becomes unstable via a vibrational mode.

In our approach we cannot exactly investigate the effect of pure
bulk viscosity on the stability of the compressible spheroids,
because the transition is influenced by the hidden
non-conservation. Taking account of bulk viscosity and violent
relaxation does not change the bifurcation point. However, in some
cases the presence of relaxation has a stabilizing influence on
the motion of the system.

We have considered an equilibrium configuration of the Maclaurin
spheroid close to the bifurcation point, in the interval of the
secular instability. After perturbation of the density, the
spheroid starts to oscillate. Our calculations have shown that in
the absence of relaxation the development of instability and the
transition of the spheroid to a three-axis object both occur. In
the presence of relaxation the perturbed spheroid remains a
spheroid and the system reaches the equilibrium state. This
behaviour takes place because during relaxation the thermal energy
increases, and the eccentricity $e$ decreases and comes into the
interval of stability. So the spheroid initially perturbed in the
interval of instability, not far from the bifurcation point,
becomes stable due to the processes of relaxation.

According hypothesis of \citet{op73}, the stability of an isolated
axially symmetric system is determined by the ratio
$U_{rot}/|U_g|$. They determined from numerical experiments the
critical value for various configuration as $0.14 \pm 0.03$.
\citet{shap04} found that compressible spheroids become secularly
unstable to triaxial deformations at the bifurcation point, where
$U_{rot}/|U_g| = 0.1375$, independent of $n$. Our formula gives
exactly the same result, which is also confirmed by our numerical
simulations. As noted above, this result remains valid also in the
presence of dissipation.

\section{Conclusions}

We have investigated the dynamics of a three-axis dark matter
ellipsoid. The equations of motion for the axes of a uniform
compressible ellipsoid have been obtained by variation of the
Lagrange function, in which violent relaxation and losses of
matter, energy and angular momentum have been included
phenomenologically.

The system was solved numerically, until the formation of
stationary rotating figures in the presence of relaxation. For
lower angular momentum $M$ we have the formation of a compressed
spheroid, while at larger $M$ we follow the development of a
three-axial instability and the formation of a three-axial
ellipsoid. The instability in this approximation happens at the
bifurcation point of the sequence of Maclaurin spheroids, where
the Jacobi ellipsoidal system starts.

The bifurcation point coinciding with the point of loss of
stability is found analytically in the form of a simple formula,
by static and dynamic approaches. Numerical and analytical
considerations give identical results.

The development of instability, connected with radial orbits, is
obtained for slowly rotating collapsing bodies.

\section*{Acknowledgments}
We are grateful to the anonymous referee for useful comments and
references.

This work is partially supported by RFBR Grant 05-02-17697-a, and
the program of RAS `Non-stationary astronomic processes'. The work
of OYuT is supported by the Dynasty Foundation.

\bsp

\label{lastpage}

\end{document}